\documentclass[useAMS,usenatbib]{mn2e}
\usepackage{graphicx}
\usepackage{natbib}
\usepackage{amsmath}
\usepackage{enumitem}
\usepackage{multirow}
\usepackage{placeins}

\makeatletter
\renewcommand*{\@fnsymbol}[1]{\ensuremath{\ifcase#1\or ^{\dagger}\or ^{*}\or \ddagger\or
   \mathsection\or \mathparagraph\or \|\or **\or \dagger\dagger
   \or \ddagger\ddagger \else\@ctrerr\fi}}
\makeatother

\newcommand{\apj}{ApJ}
\newcommand{\apjl}{ApJL}
\newcommand{\apjs}{ApJS}
\newcommand{\aj}{AJ}
\newcommand{\aap}{A\&A}
\newcommand{\mnras}{MNRAS}
\newcommand{\nat}{Nature}
\newcommand{\na}{Nature}

\newcommand{\apss}{Ap\&SS}

\newcommand{\araa}{ARA\&A}
\newcommand{\nar}{NAR}
\newcommand{\pasj}{PASJ}
\newcommand{\ssr}{SSRv}
\newcommand{\actaa}{AcA}
\newcommand{\aaps}{A\&AS}

\title[The face-on disk of MAXI~J1836$-$194]{The face-on disk of MAXI~J1836$-$194\thanks{Based on observations collected at the European Southern Observatory, Chile, under ESO Programme IDs 087.D-0914 and 089.D-0970.}}
\author[Russell et al.]{T. D. Russell$^{1}$\thanks{Corresponding email: thomas.russell@postgrad.curtin.edu.au}, R. Soria$^{1}$, C. Motch$^2$, M. W. Pakull$^2$, M. A. P. Torres$^{3,4}$, \newauthor P. A. Curran$^{1}$, P. G. Jonker$^{3,4,5}$ and J. C. A. Miller-Jones$^1$ \\
$^1$International Centre for Radio Astronomy Research - Curtin University, GPO Box U1987, Perth, WA 6845, Australia \\
$^2$Observatoire astronomique de Strasbourg, Universit\'{e} de Strasbourg, CNRS, UMR 7550, 11 rue de l'Universit\'{e}, F-67000 Strasbourg, France \\
$^3$SRON, Netherlands Institute for Space Research, Sorbonnelaan 2, NL-3584 CA Utrecht, the Netherlands \\
$^4$Harvard-Smithsonian Center for Astrophysics, 60 Garden Street, Cambridge, MA 02138, USA \\
$^5$Department of Astrophysics/IMAPP, Radboud University Nijmegen, PO Box 9010, NL-6500 GL Nijmegen, the Netherlands \\}

\begin{document}

\date{Submitted 2013 November 14}

\pagerange{\pageref{firstpage}--\pageref{lastpage}} \pubyear{2013}

\maketitle

\label{firstpage}

\begin{abstract}
We present Very Large Telescope optical spectra of the black hole candidate X-ray binary MAXI~J1836$-$194 at the onset of its 2011 outburst. Although the spectrum was taken at the beginning of the outburst and contains a significant contribution from the optically-thin synchrotron emission that originates in the radio jet, we find that the accretion disk was already large and bright. Single-peaked, narrow $\rm H\alpha$ and He {\footnotesize {II}} $\lambda 4686$ lines imply the most face-on accretion disk observed in a black hole low-mass X-ray binary to date, with an inclination angle between $4^{\circ}$ and $15^{\circ}$, assuming a black hole mass of between 5 M$_\odot$ and 12 M$_\odot$, for distances of between 4 and 10 kpc. We use New Technology Telescope observations of the system in quiescence to place strong upper limits on the mass and radius of the donor star and the orbital period. The donor is a main sequence star with a mass $\rm < 0.65 \, M_{\odot} $ and a radius $\rm < 0.59 \, R_{\odot}$ with an orbital period of $<$ 4.9 hours. From those values and Roche lobe geometry constraints we find that the compact object must be $>$1.9 M$_{\odot}$ if the system is located 4 kpc away and $>$7.0 M$_{\odot}$ at 10 kpc.

\end{abstract}


\section{Introduction}

Low-mass X-ray binaries (LMXBs) spend the majority of their lifetimes in a state of quiescence with faint or undetectable X-ray emission and an optical spectrum dominated by the companion star. Occasionally, they show dramatic X-ray brightening (outburst), the consequence of an increased accretion rate onto the central compact object; this may be due to thermal-viscous disk instabilities \citep{2001NewAR..45..449L} or an increased mass transfer rate from the donor star. An X-ray binary may go through a number of X-ray spectral states during an outburst \citep{2010LNP...794.....B,2006ARA&A..44...49R, 2004MNRAS.355.1105F}, before returning to quiescence after a few weeks or months. In the simplest classification, the two main outburst states are a hard (non-thermal) state, where most of the accretion power is liberated in a hot corona or a relativistic outflow, and a soft (thermal) state, when most of the accretion power is efficiently radiated by an accretion disk.

The main sources of energy in the disk are viscous heating and the interception and thermalisation of X-ray photons from regions close to the compact object \citep{1995xrbi.nasa...58V}. The latter generally dominates at large disk radii, where most of the optical/ultraviolet (UV) emission comes from. In the viscous-dominated part of the disk, the emitted spectrum is a multicolour blackbody with an effective temperature, $T_{\rm eff}$, scaling with radius, R, as, $T_{\rm eff}(R) \sim R^{-3/4}$ \citep{1973A&A....24..337S}, while in the irradiation-dominated outer disk, $T_{\rm eff}(R) \sim R^{-1/2}$ \citep{2002apa..book.....F}. A typical signature of external X-ray irradiation in the outer disk is the formation of a temperature-inversion layer in the disk chromosphere; as a result, He {\footnotesize {II}} $\lambda4686$ and H$\alpha$ are detected in emission, and higher lines of the Balmer series generally have narrow emission cores superposed on broader absorption formed in deeper layers \citep{1995xrbi.nasa...58V}. 

In the soft state, the optical/near-infrared (NIR) continuum comes almost entirely from the accretion disk, which becomes much brighter than the donor star \citep{1995xrbi.nasa...58V}. In the hard state there may also be a significant NIR/optical/UV contribution from the optically-thin synchrotron emission of the jet \citep{2001ApJ...554L.181J, 2002ApJ...573L..35C, 2006MNRAS.371.1334R} as well as the donor star. While X-ray studies of an X-ray binary probe the accretion processes in the region immediately surrounding the compact object, optical/UV studies provide important constraints on the size and orbital period of the binary system, and on the mass of the donor star (see \citealt{1995xrbi.nasa...58V} and \citealt{2003astro.ph..8020C} for a full review of optical and UV emission from X-ray binaries).

In this paper we present the results of an optical spectroscopic study of the X-ray transient MAXI~J1836$-$194. This LMXB was discovered in the early stages of its outburst on 2011 August 30 \citep{2011ATel.3611....1N} by MAXI/GSC on the International Space Station \citep{2009PASJ...61..999M}. The system was classified as a rapidly spinning \citep[$a=0.88\pm0.03$;][]{2012ApJ...751...34R} black hole (BH) candidate by virtue of its radio/X-ray properties \citep{2011ATel.3618....1S,2011ATel.3628....1M,2011ATel.3689....1R}. MAXI~J1836$-$194 underwent a state transition from the hard state to the hard-intermediate state (HIMS) on 2011 September 10 (inferred from INTEGRAL, {\it Swift} and RXTE X-ray spectral and timing properties; \citealt{ferrignoetal2012}). Instead of softening further to a full soft state (as is usually the case for other BH LMXBs), the outburst `failed' \citep{2004NewA....9..249B} and the source transitioned back to the hard state on 2011 September 28 and faded towards quiescence \citep{ferrignoetal2012} over several months while the jet remained active \citep{2013ApJ...768L..35R}. A re-brightening of the source was detected with \textit{Swift}/BAT in 2012 March which lasted for several months \citep{2012ATel.3966....1K, 2012ATel.3975....1Y} before a final return to quiescence by 2012 July \citep{2012ATel.4255....1Y}. From the X-ray spectral state evolution and inner-disk temperature, we argue (in our companion paper T. D. Russell et al., submitted, hereafter TDR13) that the luminosity in the early outburst phase was maximally a few per cent of the Eddington luminosity (typical for a HIMS to hard state transition; \citealt{2003A&A...409..697M,2010MNRAS.403...61D}). This suggests a plausible range of distances between 4 and 10 kpc for this source. Here, we expand on the preliminary results presented by \citet{2011ATel.3640....1P}, based on their Very Large Telescope (VLT) observations. We show that the optical/UV signatures of an accretion disk were already present in the first few days after the discovery of the outburst although the optical/NIR band contained a significant contribution from the jet \citep{2013ApJ...768L..35R}. We use these signatures to infer the inclination of the accretion disk. {\it Swift} UltraViolet/Optical Telescope (UVOT) and European Southern Observatory New Technology Telescope (NTT) measurements of the optical luminosity in quiescence also allow us to constrain the size of the donor star and hence the orbital period of the binary system, as well as place a lower limit on the black hole mass.

\section[]{Observations and reductions}

\subsection{VLT}
We observed MAXI~J1836$-$194 with the FOcal Reducer and low dispersion Spectrograph (FORS2) on the VLT as a target-of-opportunity during the implementation of Program 087.D-0870 (PI: M. Pakull). The target was observed on 2011 August 31, between Modified Julian Date (MJD) 55804.99--55805.05, and 2011 September 01 (MJD 55805.99--55806.02). We used the 1200B grism, covering the wavelength range from 3729 {\AA} to 5200 {\AA} (`blue spectra'), and the 1200R grism, from 5869 {\AA} to 7308 {\AA} (`red spectra'). The first night consisted of four 900 s observations for the blue spectra and four 450 s observations for the red spectra, while the second night consisted of two 900 s observations and two 450 s observations for the blue and red spectra respectively.

Calibration data (bias, flatfield and wavelength calibrators) were acquired following the standard calibration plan, implemented at the beginning or the end of the observing nights. We used the same slit width (1$\arcsec$.0) for all observations, giving a spectral resolution of 2.3 {\AA} for the red spectra and 1.9 {\AA} for the blue spectra on the first night and on the second night 1.6 {\AA} for both the red and the blue spectra. The orientation of the slit was at a parallactic angle of $112.5^{\circ}$ the first night, and $141.3^{\circ}$ the second night. We took $B$-band sky and through-slit acquisition images each night before starting our spectroscopic observations to confirm that the target was well centred in the slit. The average full width at half maximum (FWHM) seeing, estimated at 5000~{\AA}, was 1$\arcsec$.18 for the blue spectra and 1$\arcsec$.51 for the red spectra on the first night and 0$\arcsec$.81 and 0$\arcsec$.68 respectively on the second night.

The data were reduced with {\small ESOREX} version 3.9.6 and the FORS2 kit recipes version 4.8.7 \citep{1998Msngr..94....1A} was used to correct raw frames for bias, apply flat-fields and to apply wavelength calibration. Arc images of He, Hg and Cd, as well as Hg, Cd, Ar and Ne lamps provided wavelength solutions for the 1200B and 1200R grisms respectively, yielding a wavelength accuracy of $\sim$ 0.05{\AA} for both the blue and red spectra (taken from the scatter of the fit to all the arc lines). One-dimensional spectra were extracted from the wavelength calibrated image using a {\small MIDAS} \citep{1988igbo.conf..431B} procedure optimised for providing the best signal to noise ratio. We then used the {\small IRAF} \citep{1986SPIE..627..733T} tasks {\tt rvcorrect} and {\tt dopcorrect} to correct the spectra for the relative radial velocity of the Earth, shifting the wavelength scale to the heliocentric system. We also used standard {\small {IRAF}} tasks to combine and/or compare the spectra from the two nights. We then normalised the spectrum to the continuum with an 8$\rm ^{th}$ order cubic spline, where we masked the Balmer lines (including their broad absorption components), the interstellar lines, He {\footnotesize {II}} $\lambda 4686$ and the Bowen blend. We then measured the main properties of the lines (central position, full-width at half-maximum, equivalent width) with the {\small IRAF} task {\tt SPLOT}, where the errors are taken as the statistical error in the Gaussian fit and verified the results with the fitting package {\small QDP} \citep{tennant1991}.

\subsection{Optical/UV \textit{Swift}}
\textit{Swift} monitored MAXI~J1836$-$194 every few days during its 2011 outburst \citep{ferrignoetal2012} and 2012 re-flaring. Observations in six optical/UV filters were made with the UVOT \citep{2005SSRv..120...95R}, of which we use the three optical filters ({\it u}, {\it b} and {\it v}) that provide the strictest constraints on the luminosity of the low-mass donor star. UVOT data were pre-processed at the \textit{Swift} Data Centre \citep{2010MNRAS.406.1687B} and reduced using standard \textit{Swift} pipelines. The {\small FTOOLS}\footnote{http://heasarc.gsfc.nasa.gov/ftools/} \citep{1995ASPC...77..367B} task, {\tt uvotimsum}, was used to combine all images from the same filter. We used {\tt uvotsource} and a suitable background region to extract an aperture photometry magnitude from a source region of 2$\arcsec$.5 radius, centered on the UVOT position. Magnitudes are based on the UVOT photometric system, which differs from the Bessell system by $V-v$ $<$ 0.04 mag (for all colours) and $B-b$ $<$ 0.04 mag (for colours $b-v < 1.5$ mag; \citealt{2008MNRAS.383..627P}). The differences between the UVOT and Bessell photometric magnitudes are small compared to the uncertainty in our photometry (see Section~\ref{sec:photometry}) and we therefore treat the UVOT $v$ and $b$ magnitudes as standard $V$ and $B$ magnitudes.

\subsection{NTT}
\label{sec:ntt}
We obtained optical images during quiescence using the ESO Faint Object Spectrograph and Camera (EFOSC2; \citealt{1984Msngr..38....9B}) mounted on the NTT at La Silla Observatory, Chile. On 2013 June 11 three observations were acquired in the Gunn {\it i}-band with an exposure time of 300 s and a FWHM seeing of 1$\arcsec$.0-1$\arcsec$.1.  On 2013 June 12 and 13 single exposures of 300 s were obtained in the Gunn {\it g}, {\it r} and {\it i}-band filters with a seeing in the {\it i}-band of 1$\arcsec$.1 and 0$\arcsec$.7, respectively. Sky conditions were not photometric on all three nights, so additional 10, 30 and 60 s exposures of the field were obtained on June 12 and 13 to enable absolute flux calibration of the frames relative to data from the AAVSO Photometric All-Sky Survey data release 7 (APASS; \citealt{2009AAS...21440702H}). We used $2\times2$ on-chip binning during the observations, providing a scale of approximately 0$\arcsec$.24 per pixel. Frames were bias subtracted and then flat-fielded using dome flat-field observations. Given the faintness of the optical sources under study (see Figure~\ref{fig:optical_image}, bottom panel), we excluded data with image quality worse than 1$\arcsec$.0 FWHM seeing. 

We applied point-spread function fitting on the images using {\tt daophot} in {\small {IRAF}} to compute the instrumental magnitudes of the detected stars. Flux calibration of the field was performed using APASS objects in the MAXI~J1836$-$194 field, achieving zero-point errors \textless{} 0.1 mag. Differential photometry was performed to derive the fluxes of the sources of interest. The atmospheric calibration of the frames was defined by fitting for the reference point position, the scale and the position angle. As a first step, we obtained an astrometric solution on a 10 second {\it i}-band exposure. By comparing the positions of stars against entries from the fourth U.S. Naval Observatory CCD Astrograph Catalogue \citep[UCAC4;][]{2013AJ....145...44Z}, we found nine suitable UCAC4 sources well distributed on the image that were not saturated or blended and appear stellar. We obtained a solution with 0.0788 arcsecond RMS residuals. We then used the sources from this 10 s {\it i}-band image as secondary astrometric calibrators for our 300 s images. Taking into account uncertainties (including the systematic 0$\arcsec$.05 uncertainty that is inherent to the UCAC4 catalogue) we estimated a positional accuracy of 0$\arcsec$.13. Additionally a publicly-available 30 s {\it i}-band image of the field obtained during outburst with the Faulkes telescope on 2011 September 15 was calibrated using the same secondary reference tiers as for the 300 s NTT images. This frame was acquired with a 0$\arcsec$.3 per pixel scale under a seeing of 1$\arcsec$.2 FWHM and shows the optical counterpart at a J2000 position of R.A = 18:35:43.451, DEC = $-$19:19:10.43.

\section[]{Results}
\subsection{Optical photometry}
\label{sec:photometry}

\begin{figure}
\centering
\includegraphics[width=0.4\textwidth]{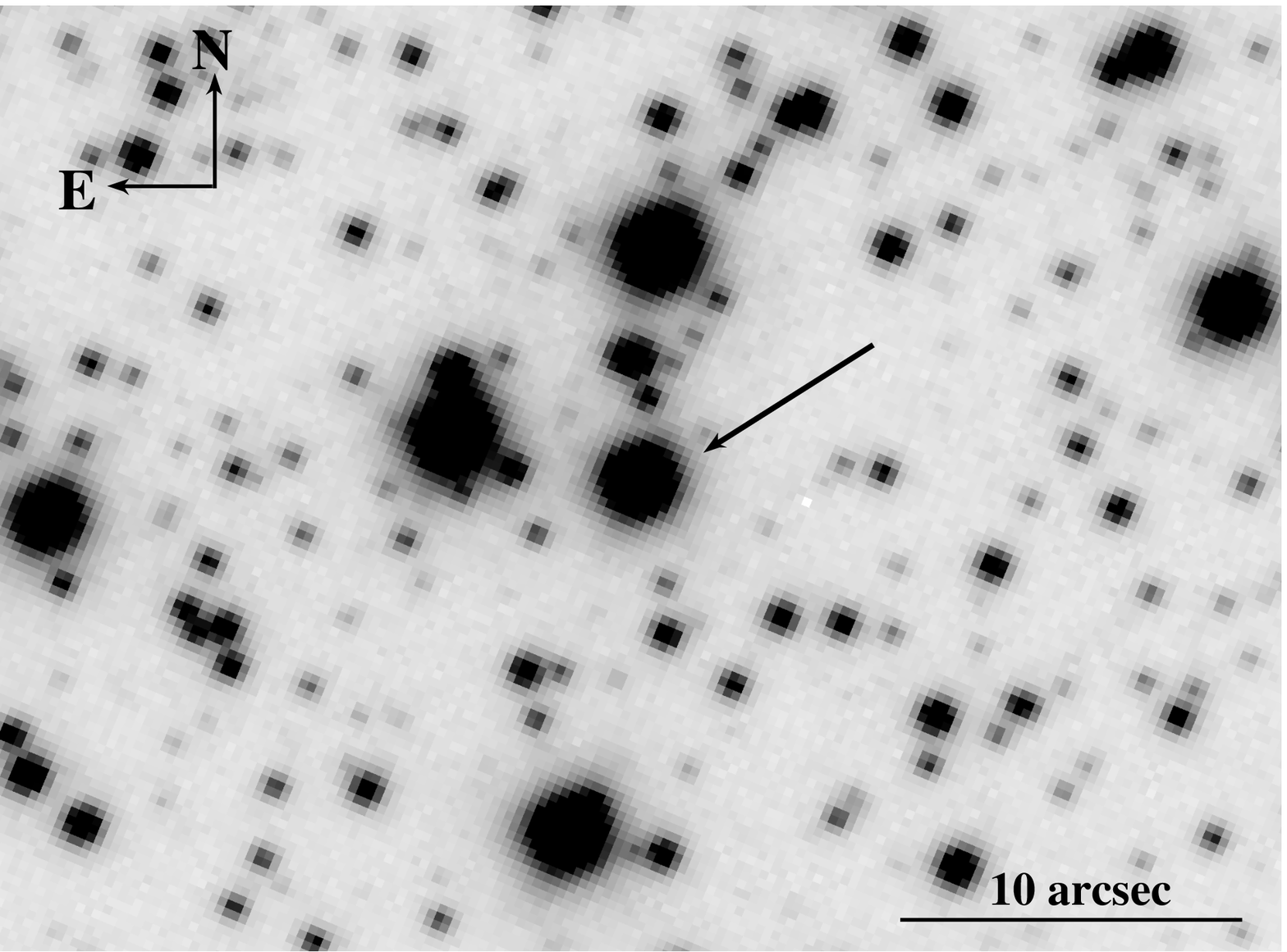}
\includegraphics[width=0.4\textwidth]{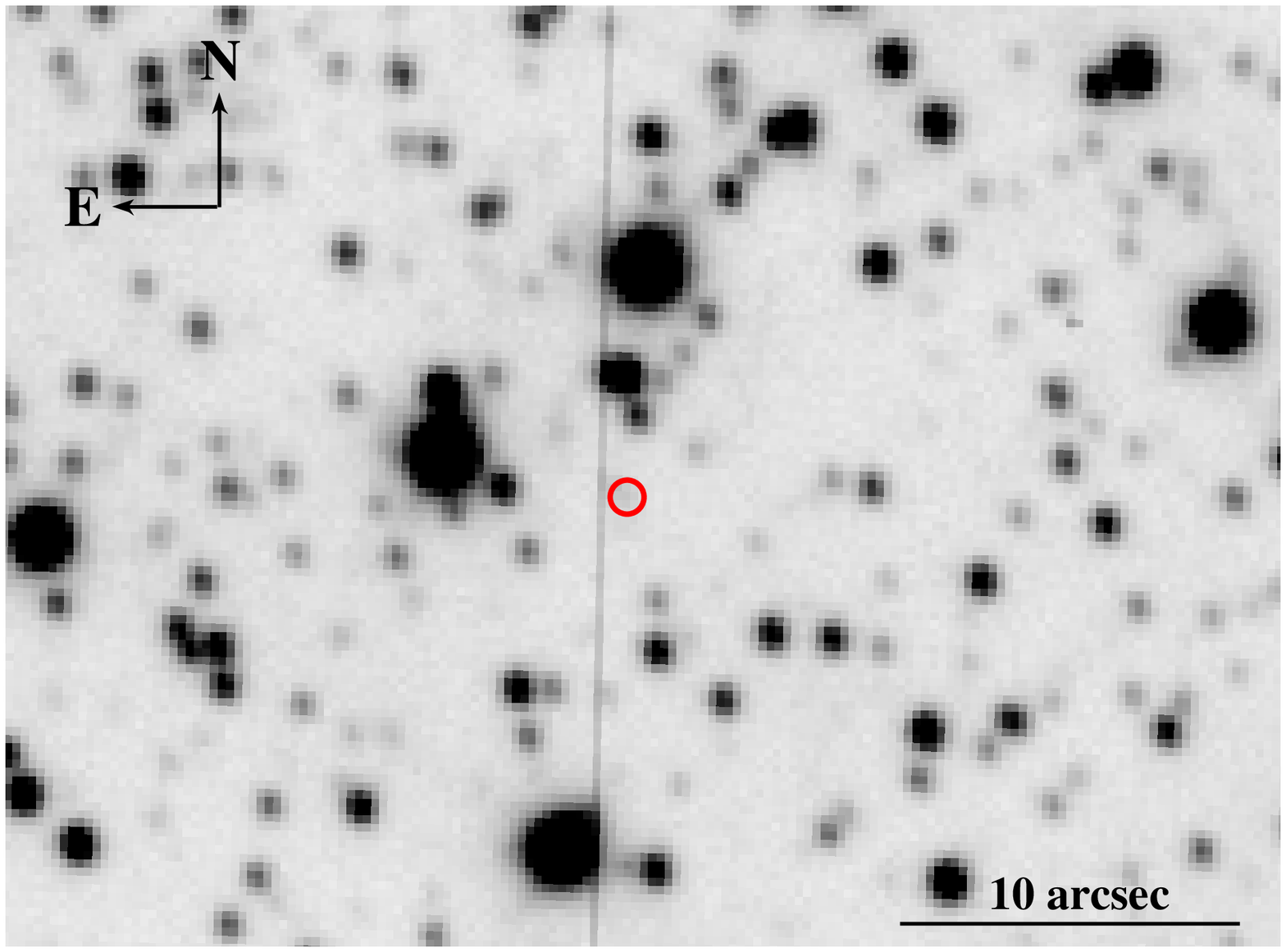}
\includegraphics[width=0.4\textwidth]{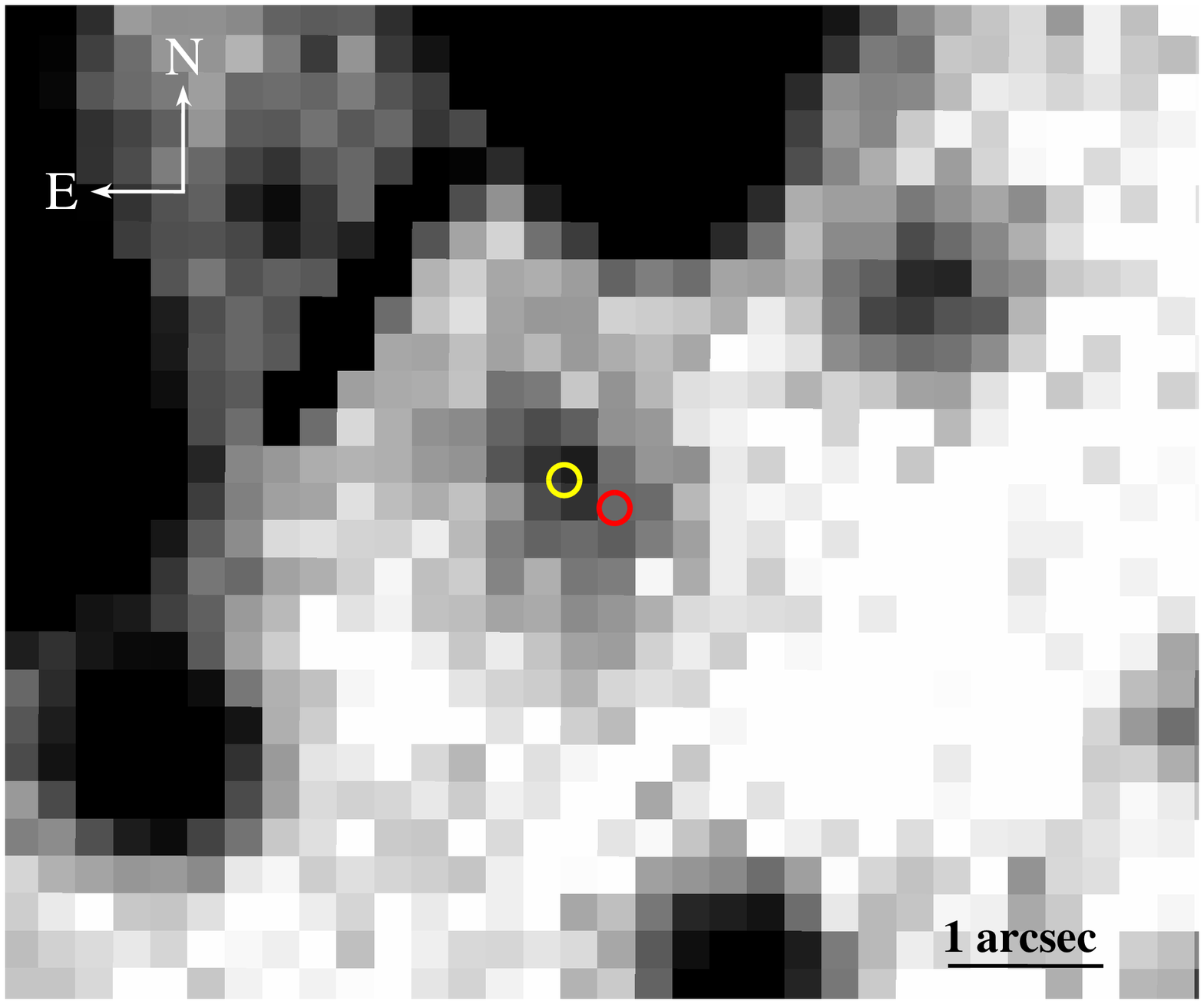}
\caption{Top panel: $B$-band image of MAXIJ1836$-$194 (marked by an arrow) and surrounding field, taken with VLT/FORS2 on 2011 August 31 as an acquisition image before our spectral observations. Middle panel: {\it i}-band NTT image of the same field taken on 2013 June 13, after the system had returned to quiescence, where the red circle of radius 0$\arcsec$.5 indicates the outburst position of the system. Bottom panel: Another {\it i}-band NTT image of the field taken on 2013 June 11 of the system in quiescence. Here the yellow circle of radius 0$\arcsec$.08 shows the unrelated field star at a J2000 position of R.A.= 18:35:43.474 and DEC= $-$19:19:10.25, while the red circle of radius 0$\arcsec$.08 shows the fitted location of the optical counterpart detected during outburst, at the J2000 position of R.A.= 18:35:43.451 and DEC = $-$19:19:10.43. The optical counterpart of MAXI~J1836$-$194 is no longer detected.}
\label{fig:optical_image}
\end{figure}

\begin{figure}
\centering
\includegraphics[width=0.47\textwidth]{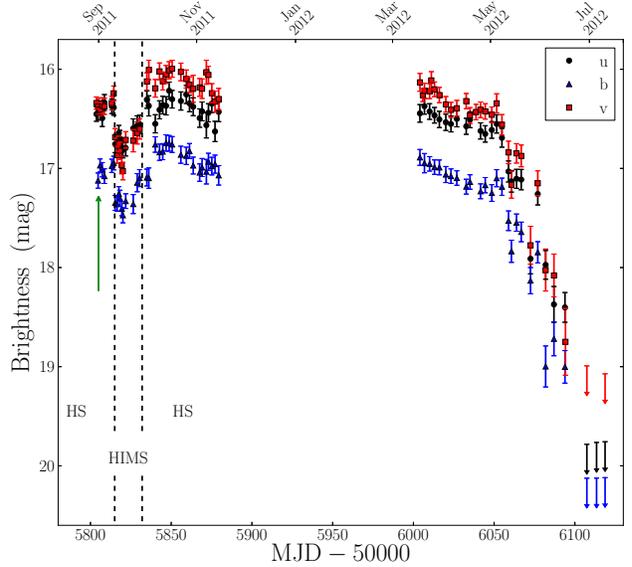}
\caption{{\it Swift}/UVOT lightcurve, showing the 2011 outburst and decline towards quiescence. The date of the first VLT observation is marked by the green arrow. The state transitions are marked by the vertical dashed lines, HS denotes the hard state and HIMS is the hard intermediate state (the states are defined by X-ray spectral and timing properties presented by \citet{ferrignoetal2012}).}
\label{fig:uvot_lightcurve}
\end{figure}

During the two nights of VLT observations (2011 August 31, 2011 September 01), MAXI~J1836$-$194 was detected as one of the brightest optical sources in the field (Figure~\ref{fig:optical_image}, top panel). Quasi-simultaneous {\it Swift}/UVOT observations on the same nights provide an accurate measurement of its optical/UV brightness and colours: we obtain $u = 16.43 \pm 0.08$ mag, $b = 17.12 \pm 0.08$ mag, $v = 16.37 \pm 0.08$ mag on the first night, and $u = 16.39 \pm 0.07$ mag, $b = 16.97 \pm 0.07$ mag, $v = 16.40 \pm 0.07$ mag on the second night (Figure~\ref{fig:uvot_lightcurve}).

Instead of brightening as the source moved into the HIMS, the optical/UV luminosity decreased; on 2011 September 17, we measured $u = 16.79 \pm 0.08$ mag, $b = 17.33 \pm 0.07$ mag, $v = 17.18 \pm 0.07$ mag from {\it Swift}/UVOT (Figure~\ref{fig:uvot_lightcurve}). The optical/UV dimming is discussed further in Section ~\ref{sec:discussion}. We used the {\it Swift}/XRT data where the soft disk-like component to the XRT spectra was largest (mid-September; TDR13) to infer the most important disk parameters; in particular, in the context of this work, the outer disk radius and X-ray reprocessing fraction. 

MAXI~J1836$-$194 underwent a re-brightening during March 2012, followed by a decline to quiescence \citep{2012ATel.3966....1K, 2012ATel.3975....1Y}. In the last three {\it Swift}/UVOT observations of 2012 June 29, July 5 and July 10, the optical counterpart was not visible, nor was it found in the stacked image from the three datasets. We take the detection limit for that combined image as the best-available 3$\sigma$ upper limit to the quiescent optical flux in the {\it u}, {\it b} and {\it v} bands: $u > 20.5$ mag, $b > 20.8$ mag, $v > 19.5$ mag (Table~\ref{tab:magnitude_limits}).

Our more recent NTT images of the system in quiescence (Figure~\ref{fig:optical_image}, bottom panel) show a number of blended sources near the location of MAXI~J1836$-$194. The brightest of these sources is a point-like object at a J2000 position of R.A.= 18:35:43.474 DEC= $-$19:19:10.25. Point spread function (PSF) photometry yields $i= 21.4\pm0.1$ mag (2013~June~11) and $g=23.7\pm0.2$ mag, $r=22.4\pm0.1$ mag and $i=21.6\pm0.2$ mag (2013~June~13). This source is 0$\arcsec$.37 away (0$\arcsec$.32 in R.A. and 0$\arcsec$.18 in DEC) from the position derived for the optical counterpart to MAXI~J1836$-$194 during outburst (R.A = 18:35:43.451 and DEC = $-$19:19:10.43, J2000; Section~\ref{sec:ntt}). Given that the images were astrometrically calibrated using the same reference stars, the uncertainty in the separation between the two objects is dominated by the uncertainty in the determination of the target profile centre. Using different algorithms implemented in {\small IRAF} to calculate these positions (centroid, Gaussian and PSF fitting), we derive positional uncertainties of 0$\arcsec$.06 and 0$\arcsec$.08 in R.A. and DEC, respectively, for the source found in the NTT images. The positional uncertainties are negligible for the bright optical counterpart detected in outburst. Therefore, the separation between the two objects is significant (5$\sigma$ in R.A.), showing that the object in the NTT images is a field star unrelated with the transient. South-West from this source there is evidence for unresolved emission overlapping the expected position for the transient. We derive 3$\sigma$ upper limits of $g> 24.0$ mag, $r> 23.7$ mag and $i> 23.5$ mag at this position for the best quality images (Table~\ref{tab:magnitude_limits}).

\begin{table}
\caption{Upper limits on the magnitude and magnitude minus the distance modulus (dm) of MAXI~J1836$-$194 during quiescence; {\it u}, {\it b} and {\it v} upper limits are determined from the summed {\it Swift}/UVOT images of 2012 June 29, July 05 and July 10, while the {\it g}, {\it r} and {\it i} upper limits are from the 2013 June 12 and 13 NTT images of the system in quiescence. The distance ranges presented cover the extent of plausible distances to the source (derived from the temperature and flux of the disk; TDR13) and magnitudes presented are not corrected for extinction.}
\label{tab:magnitude_limits}
\renewcommand{\arraystretch}{1.3}
\centering
\begin{tabular}{cccccc}
\hline
\hline
Filter & Magnitude  & \multicolumn{4}{c}{Mag$-$dm}  \\
       &            & 4 kpc & 6 kpc & 8 kpc & 10 kpc \\
\hline
\hline
{\it u} &$ > 20.5$ & $> 7.5$ & $> 6.6$ & $> 6.0$ & $> 5.5$\\
{\it b} &$ > 20.8$ & $> 7.8$ & $> 6.9$ & $> 6.3$ & $> 5.8$\\
{\it v} & $> 19.5$ & $> 6.5$ & $> 5.6$ & $> 5.0$ & $> 4.5$\\
{\it g} & $> 24.0$ & $> 11.0$ & $> 10.1$ & $> 9.5$ & $> 9.0$\\
{\it r} & $> 23.7$ & $> 10.7$ & $> 9.8$ & $> 9.2$ & $> 8.7$\\
{\it i} & $> 23.5$ & $> 10.5$ & $> 9.6$ & $> 9.0$ & $> 8.5$\\
\hline
\end{tabular}
\end{table}

\subsection{Optical spectroscopy}
\label{sec:optical}
\begin{figure}
\centering
\includegraphics[width=0.45\textwidth]{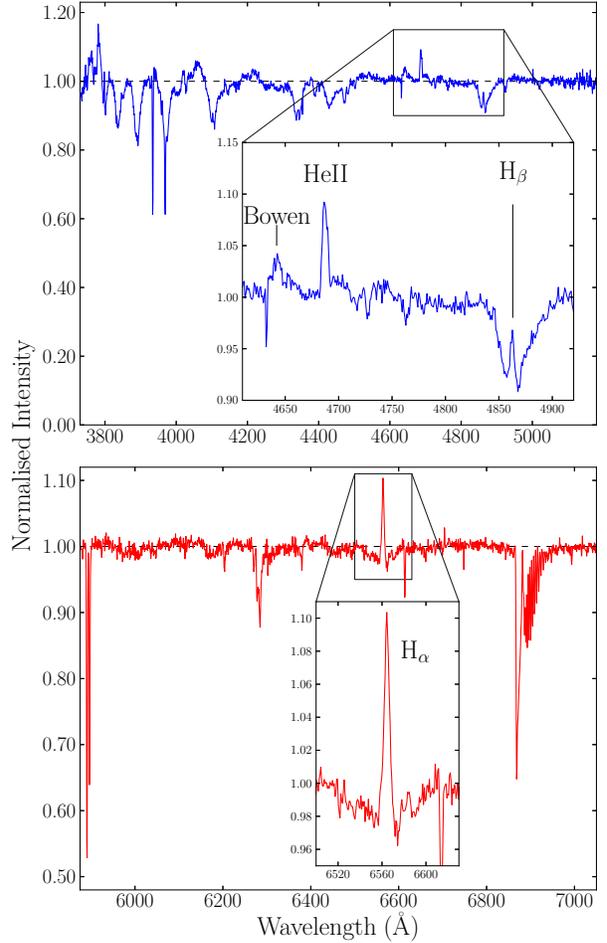}
\caption{Top panel: averaged blue spectrum of MAXI~J1836$-$194 obtained 2011 August 31 and 2011 September 01, with a zoomed portion showing the He II $\lambda4686$, H$\beta$ and Bowen region. Bottom panel: averaged red spectrum from August 31 and September 01, with a zoom of the H$\alpha$ peak. Continuum intensities have been normalised at one.}
\label{fig:spectra}
\end{figure}

Our early outburst VLT spectra (Figure~\ref{fig:spectra}) show clear signs of an optically-thick, X-ray irradiated accretion disk; rotationally-broadened He {\footnotesize {II}} $\lambda 4686$, H$\alpha$, H$\beta$ and H$\gamma$ emission in a broad absorption trough, the Bowen region at $\lambda \sim 4630$ -- $4640$ \AA, as well as other rotationally-broadened H {\footnotesize {I}} absorption lines from the Balmer series (identifiable up to $\rm H_{10}$) and He {\footnotesize {I}} absorption lines, consistent with findings presented by \citet{2011ATel.3640....1P}. 

We fitted Gaussian profiles to a selection of the de-blended Balmer emission and absorption components (Table~\ref{tab:line_width}). We found equivalent widths (EW) of the emission components, measured from the two-night combined spectrum, to be: H$\alpha = -0.83\pm0.02$~\AA, H$\beta$ emission core = $-0.25\pm0.03$~{\AA} and H$\gamma$ emission core = $-0.15\pm0.03$~{\AA} (where negative EWs represent emission and positive EWs represent absorption). We find EWs for the H$\beta$ and H$\gamma$ absorption components of $2.54\pm0.07$~{\AA} and $2.5\pm0.1$~{\AA}, respectively. For the broad and shallow H$\alpha$ absorption trough, the the FWHM is not well constrained due to the strong emission line, but we estimate the FWHM to be $\approx$ 50~{\AA} and the Full Width at Zero intensity (FWZI) to be $\approx$ 90~{\AA}. The de-reddened fluxes of the Balmer series and He {\footnotesize {II}} $\lambda 4686$ line are presented in Table~\ref{tab:line_width} and will be discussed further in Section~\ref{sec:discussion}. We determined the reddening, $E(B-V) = 0.53^{+0.03}_{-0.02}$ mag, by fitting an irradiated disk model to the broadband optical/UV/X-ray spectrum with {\small XSPEC} (TDR13). Diffuse interstellar bands provided an alternative estimate of the extinction. In particular, the $\lambda 4430$ interstellar band is clearly visible with a central absorption of 6.5\% corresponding to $E(B-V)$ = $0.6^{+0.2}_{-0.1}$ mag \citep{1987ApJ...316..449K}. We also get similar results from the $\lambda 6284$ interstellar band using the calibration from \citet{1994A&AS..106...39J}. Results from \citet{2006A&A...453..635M} indicate that there is an absorption layer of $E(B-V)\approx 0.5$ mag located within 2 kpc in the direction of the source, providing a lower limit to the distance to the source.

We determine the heliocentric velocity of the disk on both nights by fitting Gaussian profiles to a selection of suitable emission and absorption lines (Table~\ref{tab:systemic_velocity}) and measuring their average velocity shifts. The average recessional velocity is $\rm 68 \pm 20 \: km \: s^{-1}$ for the first night of VLT observations, $\rm 55 \pm 17 \: km \: s^{-1}$ for the second night and $\rm 61 \pm 15 \: km \: s^{-1}$ for the combined two-night spectrum. We also obtain consistent velocity shifts when we use only the sample of emission lines ($\rm 59 \pm 9 \: km \: s^{-1}$) or the sample of absorption lines ($\rm 60 \pm 16 \: km \: s^{-1}$); we thus conclude that both emission and absorption lines in Table~\ref{tab:systemic_velocity} come from the disk surface. In principle the recessional velocity may be different on each night due to the orbital phase of the system because the disk follows the orbital motion of the BH. However, from the derived system parameters (presented in Section ~\ref{sec:discussion}) we verify that the amplitude of the projected radial velocity of the BH around the centre of mass is $\sim$9 km s$^{-1}$ for a 5 M$_{\odot}$ BH and $\sim$5 km s$^{-1}$ for a 12 M$_{\odot}$ BH, so this effect is well within the uncertainties from our averaged velocities.

\begin{table}
\caption{The intrinsic full width at half maximum (FWHM), equivalent width (EW) and the de-reddened flux within the line for some notable lines at the onset of the 2011 outburst. Negative values represent emission and positive represent absorption. Both the emission core and broad absorption of $\rm H_{\beta}$ were recorded.}
\label{tab:line_width}
\renewcommand{\arraystretch}{1}
\centering
\begin{tabular}{cccc}
\hline
\hline
Line 				& FWHM  		& EW 		& Emitted flux  \\
				&			&		& $\times 10^{-15}$	\\
     				& (\AA)     		&(\AA)		& ($\rm erg\,s^{-1}\,cm^{-2}$) \\
\hline
\hline
	H$\alpha$(emis.)			    & 6.0$\pm$0.2	& -0.83$\pm$0.02 & 3.15$\pm$0.09 	\\
	H$\alpha$(abs.)				    & $\approx$ 50	& 1.20$\pm$0.05  & -- 	\\
	H$\beta$(emis.) 				& 4.0$\pm$0.4	& -0.25$\pm$0.03 & 1.5$\pm$0.2	\\
	H$\beta$(abs.) 					& 35.1$\pm$0.9	& 2.54$\pm$0.07	 & --	\\
	H$\gamma$(emis.)				& 4.2$\pm$0.7	& -0.15$\pm$0.03 & 1.0$\pm$0.2	\\
	H$\gamma$(abs.)				    & 23$\pm$2	    & 2.5$\pm$0.1    & -- 	\\
	He {\scriptsize II} $\lambda4686$& 5.5$\pm$0.3	& -0.52$\pm$0.03 & 3.2$\pm$0.2	\\
	Bowen blend 					& --	        & -0.58$\pm$0.06 & 3.9$\pm$0.4	\\

\hline
\end{tabular}
\end{table}

\begin{table}
\caption{Line positions and calculated systemic velocity for each individual night and for the two-night combined spectrum. The 1$\sigma$ uncertainties in the stronger lines (H$\alpha$, H$\beta$ and He {II} $\lambda 4686$) are $\pm 0.2$ {\AA} and $\pm 0.3$ {\AA} for all other lines.}
\label{tab:systemic_velocity}
\renewcommand{\arraystretch}{1.3}
\centering
\begin{tabular}{cccc}
\hline
\hline
Line  & Night 1   & Night 2   & Combined \\
 (\AA)       & (\AA) & (\AA) & (\AA)\\

\hline
 \hline
$\rm H\alpha$  6562.8 		& 6564.1 & 6563.7 & 6563.8\\
$\rm H\beta$  4861.3 		& 4862.4 & 4861.9 & 4862.3\\
$\rm H\gamma$  4340.5		& 4341.2 & 4341.3 & 4341.3\\
$\rm H\delta$  4101.7 		& 4102.0 & 4102.1 & 4102.1\\
$\rm H_{8}$	 3889.1 	    & 3890.3 & 3889.7 & 3889.7\\
$\rm H_{9}$  3835.4 		& 3836.4 & 3836.2 & 3836.3\\
$\rm H_{10}$  3799 		    & 3800.1 & 3799.7 & 3800.0\\
 He I  4026.2 			    & 4027.0 & 4027.3 & 4027.0\\
 He I  4143.8 			    & 4145.2 & 4145.1 & 4145.0\\
 He I  4387.9 			    & 4389.0 & 4388.9 & 4389.0\\
 He I  4471.5 			    & 4472.5 & 4472.0 & 4472.5\\ 
 Mg II  4481.3			    & 4482.1 & 4482.2 & 4482.1\\
 He {\footnotesize {II}}  4686	& 4686.9 & 4686.7 & 4686.8\\
\hline
Velocity (km/s)             & $68 \pm 20 $ & $55 \pm 17 $ & $61 \pm 15$ \\
\hline
\end{tabular}
\end{table}

The H$\alpha$ and He {\footnotesize {II}} $\lambda 4686$ lines are single-peaked and narrower than typically observed in Galactic LMXBs. Therefore, instead of using the peak-to-peak separation, we use their FWHM to provide an approximation of twice the projected Keplerian velocity of the disk annulus of the emitting region. For H$\alpha$ we measure a Gaussian FWHM of 6.40$\pm$0.20 {\AA} for night one and 5.75$\pm$0.20 {\AA} for night two. The He {\footnotesize {II}} $\lambda 4686$ line profile produces FWHM~=~5.57$\pm$0.30 {\AA} on night one and FWHM~=~5.43$\pm$0.30 {\AA} on night two. We deconvolved the instrumental resolution for both the red and the blue spectra on the second night (seeing limited). Averaging the intrinsic FWHMs from both nights, we obtain a velocity of $\rm 260 \pm 6 \: km \: s^{-1}$ for H$\alpha$ and $\rm 333 \pm 14 \: km \: s^{-1}$ for He {\footnotesize {II}} $\lambda 4686$. We assume that this corresponds to approximately twice the Keplerian velocity of the disk rings that give the highest contribution to those emission lines, and we use these results to determine the disk inclination (Section ~\ref{sec:inclination}). 

\section[]{Analysis and Discussion}
\label{sec:discussion}

During the first few days after the discovery of the outburst, MAXI~J1836$-$194 was in the hard spectral state with a significant contribution to the optical/UV emission from the optically-thin synchrotron emission from the radio jet (a spectral index of $\alpha =-0.6 \pm 0.1$, where $\rm flux \; S_{\nu} \propto \nu^{\alpha}$; \citealt{2013ApJ...768L..35R}). In addition to this power-law synchrotron component, the optical emission on 2011 August 31 (night 1) and September 01 (night 2) contained a multicolour thermal component from the irradiated accretion disk. From multi-wavelength spectral fitting, we estimate that $\approx$ 55\% of the $B$-band flux, and $\approx$ 70\% of the $V$-band flux came from the synchrotron component on those dates (TDR13). Two weeks after the VLT observations, the disk became the dominant source of optical emission, not because of a dramatic increase in its flux, but because of the temporary disappearance of the synchrotron component due to jet quenching \citep{2013ApJ...768L..35R}, resulting in a reduction in the total optical/UV emission (Figure~\ref{fig:uvot_lightcurve}).

\subsection{Emission lines}
\label{sec:emission_lines}
The emission lines or cores of the Balmer series are seen clearly at H$\alpha$, H$\beta$ and H$\gamma$, with H$\beta$ and H$\gamma$ being dominated by pressure-broadened absorption troughs \citep{1980MNRAS.193..793M}. In particular, the FWHM of the H$\alpha$ and H$\beta$ absorption troughs are $\approx$50 {\AA} and $\approx$35 \AA, respectively, while the FWZI of the H$\alpha$ and H$\beta$ absorption troughs are $\approx$90 {\AA} and $\approx$75 \AA, similar to other Galactic LMXBs (e.g. GRO~1655$-$40 and XTE~J1118+480; \citealt{2000ApJ...539..445S,2002ApJ...569..423T}).

While the spectra look remarkably similar, the EW of the H$\alpha$ emission line ($-0.83\pm0.02$\,\AA) is much smaller (in absolute value) than what is usually found in Galactic BH transients early in an outburst, when typical magnitudes of the equivalent widths are $\sim$ 3--6 \AA $\,$  \citep{2009MNRAS.393.1608F}. The reason for this significantly lower EW is the optically-thin synchrotron component from the jet dominating the red part of the optical spectrum, which we shall argue in Section~\ref{sec:inclination} is due to beaming because of the face-on nature of the system. Accounting for the jet component (model presented in TDR13), we find that the EW of H$\alpha$ with respect to the disk continuum alone is in fact $-4.2\pm0.1$\,\AA, similar to other Galactic XRBs \citep{2009MNRAS.393.1608F}. Similarly, the EW of He {\footnotesize {II}} $\lambda 4686$ with respect to the total blue continuum is $-0.52\pm0.03$\,\AA, but with respect to the disk continuum alone is $-1.6\pm0.1$\,\AA. The comparable EWs with respect to the disk component imply that the conditions in the disk in terms of optical depth and irradiation are similar to other LMXBs in the early stages of outburst.

From the de-reddened fluxes we find a Balmer decrement of $f_{{\rm {H}}\alpha}/f_{{\rm {H}}\beta} = 2.1\pm0.3$, lower than the canonical value of $\approx 2.87$ valid for Case-B recombination of optically thin gas at $T \approx 10^4$\,K \citep{2006agna.book.....O}; also, the ratio of $f_{{\rm {H}}\gamma}/f_{{\rm {H}}\beta} = 0.71\pm0.2$ is slightly higher than the expected Case-B recombination value of $\approx 0.466$ \citep{2006agna.book.....O}. The most likely explanation of such a discrepancy is that the lines are emitted in a layer with high electron density and optical depth, $\tau_{{\rm {H}}\alpha} > 1$, where the collisional excitation and de-excitation becomes dominant and case-B recombination is no longer valid. From ~\citet{1980ApJS...42..351D}, the measured Balmer ratios from our spectra suggest an electron density of $\sim 10^{12}$--$10^{14}$ cm$^{-3}$, consistent with their expectations for the upper layers of an optically-thick accretion disk.

For He {\footnotesize {II}} $\lambda 4686$ we assume instead that collisional processes are not significant, at least in the outer disk. In this case the de-reddened flux of this line, $f_{\rm {He II}} = (3.2\pm0.2) \times 10^{-15}$ erg cm$^{-2}$ s$^{-1}$ $= (7.5\pm0.4) \times 10^{-4}$ $\lambda 4686$ photons cm$^{-2}$ s$^{-1}$, can be used to estimate the fraction of X-ray luminosity intercepted by the disk. Following \citet{1986Natur.322..511P}, for X-ray photoionised gas, the He {\footnotesize {II}} $\lambda 4686$ photon flux is proportional to the illuminating flux of ionising photons above 54 eV; for an optically-thick region, the scaling factor is simply the ratio between the He$^+$ recombination coefficient (summed over all levels above the ground state) and the effective recombination coefficient for the $n = 4 \rightarrow 3$ transition. This ratio varies between $\approx 4.2$ and $\approx 6.3$ for temperatures $T \approx 10,000$ to $40,000$ K \citep{2006agna.book.....O}, and implies that approximately one He {\footnotesize {II}} $\lambda 4686$ photon is emitted for every $\sim 4$--$6$ UV/X-ray photons that are intercepted by the disk. The unabsorbed, isotropic photon flux above 54 eV measured from {\it Swift} XRT data taken on August 31 is $\approx 3.9$ photons cm$^{-2}$ s$^{-1}$ (TDR13), meaning that we should observe approximately (0.65 -- 0.98) $f_{\rm out}$ $\lambda 4686$ photons cm$^{-2}$ s$^{-1}$, where $f_{\rm out}$ is the disk interception and reprocessing fraction. As $\approx 7.5 \times 10^{-4}$ $\lambda 4686$ photons cm$^{-2}$ s$^{-1}$ are observed this means that the outer accretion disk must be intercepting $\sim 10^{-3}$ of the UV/X-ray photons that are emitted from closer to the BH; this reprocessing fraction is in good agreement with theoretical models \citep{1990A&A...235..162V, 1996A&A...314..484D, 2002MNRAS.331..169H, 2009MNRAS.392.1106G}, demonstrating that, even at this early stage of the outburst, the accretion disk is already relatively large, bright and optically thick. 

\subsection{Inclination angle of the disk}
\label{sec:inclination}
The profiles and widths of the H$\alpha$ and He {\footnotesize {II}} $\lambda 4686$ emission lines are good indicators of the Keplerian rotational velocity 
of the disk annuli responsible for their emission. Instead of the usual flat-topped or double-peaked profile \citep{1977ApJ...216..822P, 1981AcA....31..395S}, we observe single-peaked and relatively narrow H$\alpha$ and He {\footnotesize {II}} $\lambda 4686$ profiles (Figure~\ref{fig:spectra}), which are indicative of a face-on accretion disk \citep{1981AcA....31..395S}. Using the FWHM provides a good approximation of twice the Keplerian velocity of the disk annulus that contributes most to the line emission\footnote{In fact, the FWHM is a slight overestimate of the projected rotational velocity \citep{1981AcA....31..395S}, but the difference is $\la 10$\% for the moderately flat radial temperature profile expected in the outer region of an irradiated disk. In the absence of more direct measurements, the FWHM of the disk lines is a robust and conservative approximation to the projected rotational velocity of the emitting gas; if anything, the true projected rotational velocity may be slightly lower, which would further reduce the inclination angle estimated here.}.

First, we determined whether the peak of the H$\alpha$ emission originates in the outermost disk annulus, that is, whether $R_{\rm{H}\alpha} \approx R_{\rm out}$. To determine $R_{\rm out}$, we modelled the broadband optical/UV/X-ray spectrum (TDR13) with the irradiated disk model {\it diskir} \citep{2008MNRAS.388..753G} within {\footnotesize XSPEC} \citep{1996ASPC..101...17A}; the main free parameters of this model are the colour temperature of the inner disk $T_{\rm in}$, the disk-blackbody normalisation constant $K$, the fraction of X-ray flux intercepted and reprocessed in the outer disk $f_{\rm out}$, and $R_{\rm out}$ expressed as a function of the inner-disk (fitting) radius $r_{\rm in}$. The X-ray data provide strong constraints on $T_{\rm in}$ and $K$ (and therefore also $r_{\rm in}$; \citealt{2007Ap&SS.311..213S}), while the optical/UV data constrain $R_{\rm out}/r_{\rm in}$ and the irradiation fraction. In particular, we find typical irradiation fractions $f \approx 10^{-3}$ (in agreement with what we estimated from the X-ray/$\lambda 4686$ flux ratio, Section~\ref{sec:emission_lines}), and typical values of $y \equiv \log \left(R_{\rm out}/r_{\rm in}\right) \approx 4.1$ (TDR13). Since $r_{\rm in} \approx D_{\rm 10 kpc} \, \left[K/\cos(i)\right]^{1/2}$, we obtain $R_{\rm out} = 10^y r_{\rm in} \approx 1.4 \times 10^{11} \cos(i)^{-1/2} D_{\rm 10 kpc}$ cm. The Rayleigh-Jeans tail of the irradiated outer disk spectrum suggests that $T(R_{\rm out}) \approx 15,000$ K. This characteristic temperature confirms that the outermost disk annulus must be a strong Balmer emitter \citep{2006agna.book.....O}. Therefore, we are justified in assuming that the projected rotational velocity of the outer disk $v_{R_{\rm out}} \approx v_{\rm H\alpha} \approx \rm 130 \pm 4 km/s$. 

Using other well-known analytic models of irradiated disks \citep{1990A&A...235..162V,1997ApJ...488...89K,1999MNRAS.303..139D}, we obtain a temperature range $\approx 10,000$--$20,000$ K for a radius $\approx 10^{11}$ cm and an X-ray luminosity $\approx 10^{37}$ erg s$^{-1}$; typical of this system and consistent with the H$\alpha$ emission. Even if we did not have any X-ray information available, we could still estimate $R_{\rm out}$ and $T(R_{\rm out})$ simply by fitting a blackbody curve to the characteristic spectral peak in the optical/UV region that is typical of irradiated disks. This approximation can be qualitatively understood from the fact that the outer disk annuli dominate the optical emission by virtue of their large emitting area and moderately flat radial temperature gradient. It can also be quantitatively verified by re-fitting a simple blackbody to the optical/UV peak of a full irradiated disk spectrum. With this simpler method, we obtain a characteristic value of $R_{\rm out} \approx 1.4 \times 10^{11} \cos(i)^{-1/2} D_{\rm 10 kpc}$ cm at a temperature $T \approx 15,000$ K, in agreement with what we obtained from our full optical to X-ray model. We can now write:

\begin{align}
v_{\rm H\alpha} & \approx v_{\rm out} \approx \sqrt{\frac{G \: M_{\rm BH}}{R_{\rm out}}} \sin(i)\notag \\
	      &\approx \frac{\left( G M_{\rm BH} \right)^{1/2}}{\left( 10^y D_{\rm 10 kpc}\right)^{1/2} K^{1/4}} \sin(i)[\cos(i)]^{1/4}, 
\label{eq:v_out}
\end{align}
where $v_{\rm H\alpha} \approx \rm 130 \pm 4$ km s$^{-1}$. The rotational broadening of the emission cores in higher Balmer lines ($v_{\rm H\beta} \approx 123$ km s$^{-1}$; $v_{\rm H\gamma} \approx 146$ km s$^{-1}$; Table~\ref{tab:line_width}) is comparable to that of H$\alpha$; however, the uncertainty is larger because H$\beta$ and H$\gamma$ are dominated by the broader absorption trough. The rotational broadening of He {\footnotesize {II}} $\lambda 4686$ is slightly higher ($v_{{\rm He II}} = 166 \pm 7$ km s$^{-1}$), consistent with peak emission at a characteristic radius $R_{{\rm He II}} \approx 0.6 R_{\rm{H}\alpha}$, at higher temperatures. We numerically solve Equation~\ref{eq:v_out} for $i$ as a function of distance and BH mass. We find that MAXI~J1836$-$194 has a face-on disk (Table~\ref{tab:rout}), with {\it i} between $\rm 4^{\circ}$ and $15^{\circ}$ (at a 90\% confidence limit) for a distance of 4 -- 10 kpc and a BH mass of between 5 -- 12 M$_\odot$ (the typical range of Galactic BH masses; \citealt{2012ApJ...757...36K}). 

The inclination angle we determined is much lower than the $\sim 20^\circ$ to $\sim 80^\circ$ range seen in other LMXBs \citep{2013MNRAS.tmp.1239M,2005ApJ...623.1017N} and could be the reason why this system is so suitable for jet and jet break studies \citep{2013ApJ...768L..35R}, as Doppler boosting would make the jet appear much brighter and may shift the jet break to higher frequencies than in similar systems with a less favourable orientation. The lack of phase-dependent modulation in the {\it Swift}/UVOT lightcurve (Figure~\ref{fig:uvot_lightcurve}) supports our conclusion of a face-on accretion disk and the flat-topped shape of the hardness-intensity diagram (\citealt{ferrignoetal2012}, figure 5) is also consistent with other relatively low-inclination LMXBs \citep{2013MNRAS.tmp.1239M}, even though the system never settled into a full soft state. We also note that the widths of the Balmer absorption troughs are similar between this face-on system and other higher-inclination systems suggesting that the broadening is mostly due to pressure effects, rather than Keplerian rotation.

\begin{table}
\caption{Outer disk radii and inclination angles for a range of plausible BH masses \citep{2012ApJ...757...36K} and source distances.}
\label{tab:rout}
\renewcommand{\arraystretch}{1.5}
\centering
\begin{tabular}{cccc}
\hline
\hline
BH mass & Distance to source & $\rm R_{out}$ & Inclination angle \\
		& (kpc) & ($\rm \times 10^{10}$) cm & ($\rm i^{\circ}$) \\
\hline
\hline
\multirow{4}{*}{5 $\rm M_{\odot}$} & 4 & $5.3^{+1.3}_{-1.6}$  & $7.0^{+2.2}_{-1.0}$ \\
 				   & 6 & $7.9^{+2.0}_{-2.3}$  & $8.5^{+2.7}_{-1.2}$ \\
 				   & 8 & $10.6^{+2.6}_{-3.1}$  & $9.8^{+3.1}_{-1.4}$ \\
 				   & 10 & $13.2^{+3.3}_{-3.9}$  & $11.0^{+3.5}_{-1.5}$ \\
\hline
\multirow{4}{*}{8 $\rm M_{\odot}$} & 4 & $5.3^{+1.3}_{-1.6}$  & $5.5^{+1.8}_{-0.8}$ \\
 				   & 6 & $7.9^{+2.0}_{-2.3}$  & $6.7^{+2.1}_{-1.0}$ \\
 				   & 8 & $10.5^{+2.6}_{-3.1}$  & $7.8^{+2.5}_{-1.1}$ \\
 				   & 10 & $13.2^{+3.3}_{-3.9}$  & $8.7^{+2.8}_{-1.2}$ \\

\hline
\multirow{4}{*}{12 $\rm M_{\odot}$} & 4 & $5.3^{+1.3}_{-1.6}$  & $4.5^{+1.4}_{-0.6}$ \\
 				   & 6 & $7.9^{+2.0}_{-2.3}$  & $5.5^{+1.7}_{-0.8}$ \\
 				   & 8 & $10.5^{+2.6}_{-3.1}$  & $6.3^{+2.0}_{-0.9}$ \\
 				   & 10 & $13.2^{+3.3}_{-3.9}$  & $7.1^{+2.3}_{-1.0}$ \\
\hline
\end{tabular}
\end{table} 

\subsection{Donor star size and orbital period} 
\begin{figure}
\centering
\includegraphics[width=0.45\textwidth]{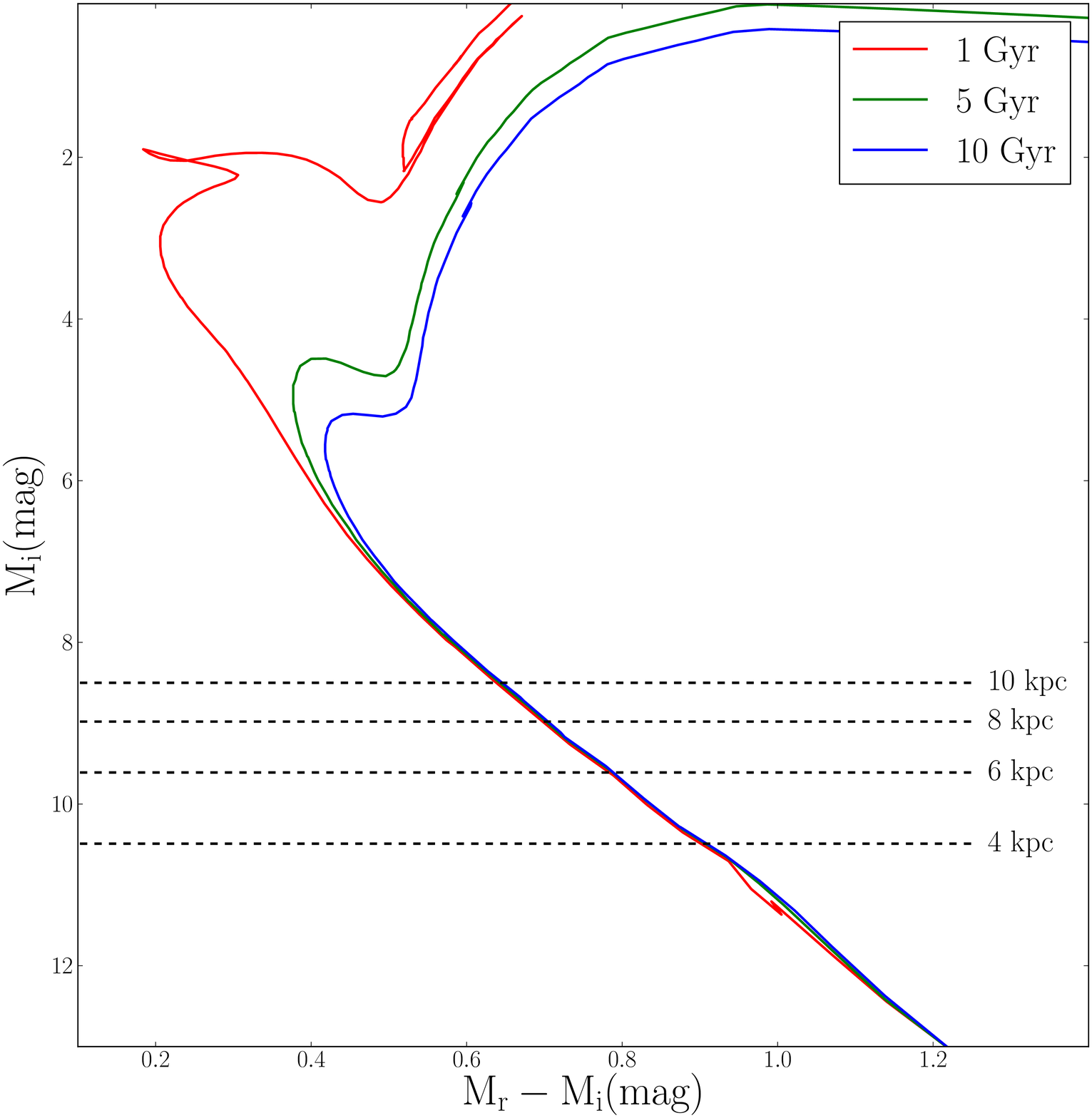}
\caption{Padova stellar evolutionary models for stellar populations of age 1 Gyr, 5 Gyr and 10 Gyr \citep{2012MNRAS.427..127B}, corrected for reddening ($E(B-V)=0.53$ mag) to compare with the magnitudes presented in Table~\ref{tab:magnitude_limits}. Coloured lines represent the stellar population of given ages and black dashed lines are the upper limits on the magnitude from quiescent photometric observations.}
\label{fig:isochrone}
\end{figure}

LMXBs are usually associated with older stellar populations \citep{2012A&A...546A..36Z}. Using solar-abundance isochrones based on the Padova stellar evolutionary models\footnote{http://stev.oapd.inaf.it/cgi-bin/cmd} \citep{2012MNRAS.427..127B} for stellar populations of age 1 Gyr, 5 Gyr and 10 Gyr, we place mass and radius constraints (Figure~\ref{fig:isochrone}) on the companion star from the deepest quiescent luminosity limits of MAXI~J1836$-$194 (Table~\ref{tab:magnitude_limits}). Generating a theoretical stellar population based on the {\it Swift}/UVOT and NTT photometric systems we find that the donor must be a main sequence star with a mass $\rm <0.65 \: M_{\odot}$ and a radius $\rm <0.59 \: R_{\odot}$ (this occurs for a 1 Gyr population at a distance of 10 kpc). The binary period, {\it P}, can be determined solely from the mean density ($\bar{\rho}$) of the Roche lobe-filling companion star \citep{2002apa..book.....F},
\begin{equation}
\bar{\rho} = \frac{3 M_{2}}{4 \pi R_{2}^{3}} \cong \frac{3^5 \pi}{8 G P^2} \cong 110 P^{-2}_{\rm hrs} \; {\rm\, g \, cm^{-3}}. \\
\label{eq:period}
\end{equation}
As the companion is on the lower main sequence, we know that its radius and mass are approximately equal in solar units \citep{1990sse..book.....K,2002apa..book.....F}, hence we place an upper limit on the binary period of $<$4.9 hours. Table~\ref{tab:total_limits} presents the maximum allowable companion mass and radius, as well as the binary period for all population ages at distances of 4, 6, 8 and 10 kpc. Our derived mass, radius and orbital period values depend on the extinction. If we consider the lowest plausible reddening of $E(B-V)=0.5$ mag (see Section~\ref{sec:optical}) we find that, for all plausible distances, the donor star mass and radius would only reduce by $<0.05$ (in M$_{\odot}$ and R$_{\odot}$, respectively), and the orbital period would decrease by a maximum of $\sim 0.05$ hours. If the reddening is increased to $E(B-V)=0.7$ mag, the limits on the mass and radius increase by $<0.05$ (in M$_{\odot}$ and R$_{\odot}$, respectively) and the orbital period will only increase by a maximum of 0.5 hours, for all plausible distances. This demonstrates that uncertainty in the extinction does not have a significant effect on our companion star results. The mass and radius limits restrict the donor star to being a K or M type main sequence star with a mass and radius that are consistent with other transient BH LMXB systems such as GRO~J0422$+$32 and XTE~J1118$+$480 \citep{2006ARA&A..44...49R}.

\begin{table}
\caption{Largest possible mass, size and period estimates derived from quiescent {\it Swift}/UVOT and NTT observations and Padova stellar evolutionary models of different population ages. Presented are the maximum limits for all population ages.}
\label{tab:total_limits}
\renewcommand{\arraystretch}{1.3}
\centering
\begin{tabular}{ccccc}
\hline
\hline
Distance & 4 kpc  & 6 kpc & 8 kpc & 10 kpc \\
\hline
$M_2$ ($\rm M_{\odot}$)   & $<0.33$ & $<0.45$ & $<0.55$ & $<0.65$ \\
$R_2$ ($\rm R_{\odot}$) & $<0.31$ & $<0.41$ & $<0.50$ & $<0.59$  \\
Period (hrs)             & $<2.7$  & $<3.4$  & $<4.1$ & $<4.9$ \\
\hline
\end{tabular}
\end{table}

\subsection{Black hole mass}
From the mass and radius limits we place on the companion star, and the outer disk radius from X-ray analysis (Table~\ref{tab:rout}), we are able to infer a lower limit to the BH mass. The size of the Roche lobe is dependent on the mass ratio ($q=M_2/M_1$) and the binary separation {\it a}, approximated by the form \citep{1983ApJ...268..368E,2002apa..book.....F}:
\begin{equation}
\frac{R_{\rm L2}}{{\it a}} \approx \frac{0.49{\it q}^{2/3}}{0.6{\it q}^{2/3}+\ln\left(1+{\it q}^{1/3}\right)}. \\
\label{eq:rl}
\end{equation}
Similarly, we obtain the average Roche lobe radius of the primary by replacing {\it q} with {\it q}$^{-1}$ in Equation ~\ref{eq:rl}. The ratio of the two radii is:
\begin{equation}
\frac{R_{\rm L2}}{R_{\rm L1}} \approx \frac{q^{4/3}\left[0.6q^{-2/3}+\ln\left(1+q^{-1/3}\right)\right]}{0.6q^{2/3}+\ln\left(1+q^{1/3}\right)} \approx q^{0.45}, \\
\label{eq:rlratio}
\end{equation}
where the second approximation holds for 0.03 $<$ q $<$ 1. In Equation ~\ref{eq:rlratio}, $M_2$ and $R_2$ (= $R_{\rm L2}$) are constrained by the photometric observations in quiescence, and $R_{\rm L1} \approx 1.3 R_{\rm out}$ \citep{1977ApJ...216..822P,1988MNRAS.232...35W} is obtained from the X-ray spectrum of the disk in outburst (as a function of distance). Thus, we can use Equation ~\ref{eq:rlratio} to place a lower limit on the mass of the primary. We find that the compact object must have a mass $M_1 > 1.9 M_{\odot}$ if the system is located at 4 kpc, $M_1 > 3.8 M_{\odot}$ at 6 kpc, $M_1 > 5.5 M_{\odot}$ at 8 kpc, and $M_1 > 7.0 M_{\odot}$ at 10 kpc. This confirms that the compact object is a stellar-mass BH, unless at the lowest projected distance (4 kpc), with a mass comparable to other transient BH LMXBs \citep{2012ApJ...757...36K}. With this same method, if we could detect and measure the quiescent magnitude of the donor star (rather than just the upper limit) with deeper photometric observations, we could infer the BH mass without the need for phase-resolved spectroscopic measurements of radial velocity shifts (which are difficult to do in face-on systems). 

\section{Conclusions}

We have used optical spectroscopic and photometric observations to determine or constrain the binary system parameters of MAXI~J1836$-$194. From the optical luminosity of the irradiated disk, combined with the relative narrowness of the disk emission-line profiles, we estimate that the inclination angle of the disk is between $\approx 4^{\circ}$ and $\approx 15^{\circ}$ for a range of distances suggested by the X-ray luminosity. This is the lowest inclination angle found in any transient BH LMXB, and is likely to produce noticeable beaming effects in our measurements of the jet emission properties, making this system ideal for studies of the spectral break evolution during outburst \citep[][TDR13]{2013ApJ...768L..35R}. 

The donor star was not detected in our observations of the source after its return to quiescence; however, we can still place strong upper limits on its luminosity and determine that it must be a main sequence K or M star. For such stars, an upper limit to the luminosity also gives upper limits to the mass and radius ($M_2 < 0.65 M_{\odot}$ and $R_2 < 0.59 R_{\odot}$), and the binary period ($P < 4.7$ hr), because the star must fill its Roche lobe for mass transfer to occur. Combining our photometric constraints of donor star mass and radius with the outer radius of the accretion disk in outburst from X-ray analysis ($R_{\rm out} \approx 1 \times 10^{11}$ cm at 8 kpc), and therefore also the average radius of the BH Roche lobe, we obtain lower limits to the primary mass. This confirms that, except possibly at the lowest distance range, the primary is a typical stellar mass BH with similar mass to other Galactic low-mass X-ray binary transients.

\section*{Acknowledgments}
We would like to thank the anonymous referee for comments that have improved this manuscript. We also thank Gary Da Costa and Daniel Bayliss for their contribution and help with optical monitoring. This research was supported under the Australian Research Council's Discovery Projects funding scheme (project number DP 120102393). This research has made use of the APASS database, located at the AAVSO web site. Funding for APASS has been provided by the Robert Martin Ayers Sciences Fund.

\label{lastpage}

\end{document}